\documentclass[aps,prl,twocolumn,floatfix,superscriptaddress]{revtex4-2} 

\usepackage{appendix}
\usepackage{hyperref}
\usepackage{graphicx}  
\usepackage{dcolumn}   
\usepackage{bm}        
\usepackage{amssymb}   
\usepackage{amsmath}
\usepackage{xcolor}
\usepackage{braket}
\usepackage[T1]{fontenc} 
\graphicspath{figures/}
\begin{document}
\def\TSD{$T_\text{SD}$}
\def\TID{$T_\text{ID}$}
\def\Si{$^{29}$Si~}

\newcommand{\fitFigScale}{0.9}

\newcommand{\sr}[1]{\textcolor{blue}{#1}}
\newcommand{\lz}[1]{\textcolor{red}{#1}}
\newcommand{\eqw}[1]{\textcolor{magenta}{#1}}

\title{Spectral diffusion of phosphorous donors in silicon at high magnetic field}

\author{Ethan Q. Williams}  \thanks{These authors contributed equally to this work.} \affiliation{Department of Physics and Astronomy, Dartmouth College, Hanover, New Hampshire 03755, USA}
\author{Lihuang Zhu}  \thanks{These authors contributed equally to this work.} \affiliation{Department of Physics and Astronomy, Dartmouth College, Hanover, New Hampshire 03755, USA}
\author{Johan van Tol} \affiliation{National High Magnetic Field Laboratory, Florida State University, Tallahassee, Florida 32310, USA} 
\author{Chandrasekhar Ramanathan} \email[]{sekhar.ramanathan@dartmouth.edu} \affiliation{Department of Physics and Astronomy, Dartmouth College, Hanover, New Hampshire 03755, USA} 

\vskip 0.25cm

\date{\today}

\begin{abstract}
We study the central spin physics of a phosphorus donor electron in silicon interacting with a silicon-29 bath at high magnetic field (8.59 T).  
We find that the spectral diffusion time is shorter and exhibits a larger anisotropy with respect to crystal orientation in the magnetic field than in previous measurements at 0.35 T. The increased anisotropy suggests a modification of the hyperfine interactions at high field. The 1.2 THz cyclotron energy is a significant fraction of the 10.8 THz Rydberg energy of the bound donor, which can result in a non-trivial magnetic perturbation of the hydrogenic donor wavefunction.
Low-power, above-bandgap optical excitation is seen to increase the spectral diffusion time, recovering the low-field spectral diffusion time at most crystal orientations.  Understanding such perturbations to the spatial wavefunction of donor electron spins could be key to engineering their high-fidelity control.


\end{abstract}

\maketitle


\noindent Donor-bound electron spins in phosphorus-doped silicon (Si:P) exhibit some of the longest coherence times observed in solid state systems at liquid helium temperatures \cite{gordon_microwave_1958, tyryshkin_electron_2003, tyryshkin_coherence_2006, tyryshkin_electron_2012,muhonen_storing_2014}. 
The scalability of manufacturing silicon devices and their compatibility with complementary metal-oxide semiconductor electronics -- which allows integrating classical controls on the same chip -- has made silicon-based quantum computers a highly sought after goal in the past two decades \cite{kane_silicon-based_1998, ogorman_silicon-based_2016, morello_donor_2020, mccallum2021donor, jones2023atomic,hsueh2024engineering,song2025entanglement}.

Phosphorus is a shallow donor in silicon with an extended electronic Bohr radius of approximately 1.8 nm \cite{smith_ab_2017}. Electron Paramagnetic Resonance (EPR) experiments on bound donor electrons need to be performed at liquid helium temperatures (or lower) to avoid donor ionization. With silicon having an atomic density of $5.0 \times 10^{22}$ atoms/cm$^3$, there are 1200 silicon nuclei contained within a sphere of the Bohr radius.
Of the three silicon isotopes present in natural abundance silicon --  $^{28}$Si, $^{29}$Si and $^{30}$Si -- only the 4.67\% abundant \Si~has non-zero spin (S= 1/2) and couples to the donor electron via both contact and dipolar hyperfine interactions.  Since the spin polarization of the \Si spins is low at laboratory magnetic fields, magnetic dipolar interactions between these spins induce energy-conserving spin flip-flops that continually change the configuration of the \Si~spins.  This leads to a fluctuating local field at the donor electron spin site, resulting in the phenomenon of spectral diffusion \cite{schweiger_principles_2001}. This is the dominant source of signal decay in spin echo EPR experiments on lightly doped Si:P samples, and has been extensively studied within the framework of the central spin problem \cite{de_sousa_electron_2003, tyryshkin_electron_2003, tyryshkin_coherence_2006, tyryshkin_electron_2012, ma_uncovering_2014, ma_classical_2015, witzel_quantum_2005, witzel_quantum_2006, witzel_decoherence_2007, witzel_electron_2010, yang_quantum_2016}.
The elimination of $^{29}$Si nuclei was seen to dramatically suppress spectral diffusion of the phosphorus donors \cite{tyryshkin_electron_2012}, and such enriched  $^{28}$Si samples are now used in most quantum information processing protocols \cite{muhonen_quantifying_2015,he_two-qubit_2019,madzik_precision_2022}.  

High magnetic fields \cite{van_tol_high-field_2009} offer an alternative way to initialize electron spins to a nearly pure state compared to operating at dilution refrigerator temperatures, and could provide improved $g$-factor resolution for selectively addressing individual spins.  Here we investigate the coherence properties of Si:P in a natural abundance silicon crystal at 8.59 T.  We find that the characteristic spectral diffusion time (\TSD) can be more than a factor of 2 shorter than the values measured at X-band (0.35 T, 9 GHz) at the same temperature and orientation. We also find that low-power above-bandgap optical excitation \textit{increases} the spectral diffusion time of the donor electron spin by a factor of two.  

Our results suggest that the \Si~hyperfine interactions are modified at high magnetic field. This is most likely due to the magnetic perturbation of the donor electron wavefunction under these conditions.  The \Si~spins become a valuable spatial probe of the donor electron wavefunction in these pulsed EPR experiments. Understanding and controlling the spatial profile of the electron wavefunction could be key to engineering high-fidelity exchange gates at higher magnetic fields.



%



\begin{figure*}[htb]
  \includegraphics[width=0.9\linewidth]{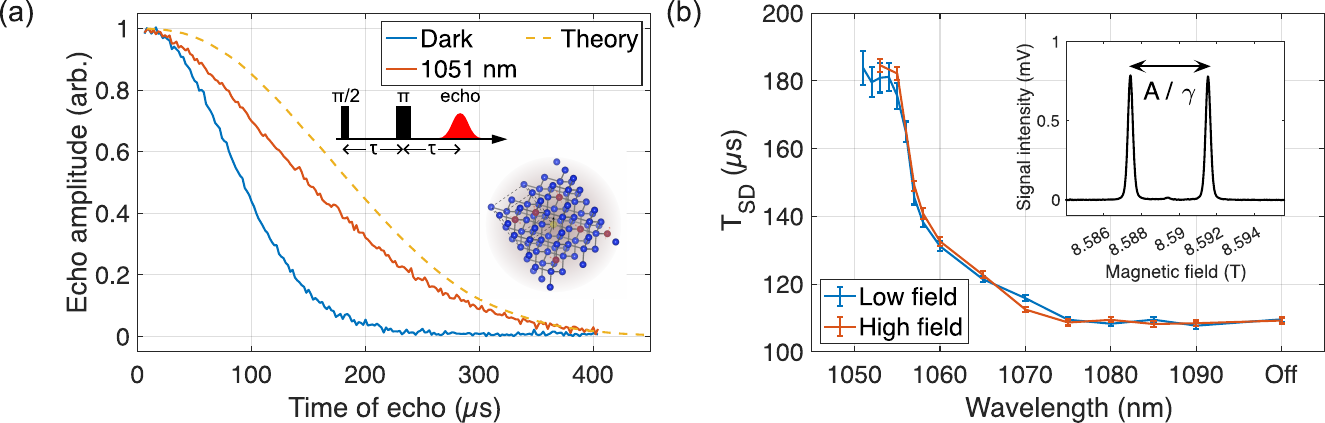}
  \vspace*{-0.1in} 
  \caption{(a)  Experimental Hahn echo decay (normalized) acquired on the 8.5915 T transition with Sample 1 (estimated to be close to the [111] or [011] orientation). The blue curve (``Dark'') is acquired without any optical excitation. The yellow curve (``Theory'') is the theoretical prediction from the pair correlation approach of Witzel \textit{et al}.\  \cite{witzel_quantum_2005} (Equation~\ref{eq:Witzel})  which agrees with low-field experiments. The red curve (``1051 nm'') shows the signal acquired under low-power ($\sim$0 dBm) 1051 nm optical excitation.  The inset shows a schematic of the Hahn echo experiment and an illustration of the silicon lattice with a central phosphorus donor.  (b) 
 Variation of the spectral diffusion time \TSD~with laser wavelength in Hahn echo experiments performed on Sample 1 at the two EPR transitions. The spectral diffusion time increases from 110(2) $\mu$s to 185(5) $\mu$s as the energy of the exciting photon is increased above the silicon bandgap (approximately $1060$ nm at 4.2 K). The inset shows the field-swept continuous wave EPR spectrum of bulk Si:P (Sample 1) at 4.3 K showing the two hyperfine-split transitions due to the $^{31}$P nuclei. } 
 \label{fig:fig1}
\end{figure*}

The experiments were performed on a lightly-doped (N$_D = 3.3-3.5 \times 10^{15}$ cm$^{-3}$) natural isotopic abundance Si:P at 4.2 K and 8.59 T using the 240 GHz electron magnetic resonance setup at the National High Magnetic Field Laboratory (NHFML) \cite{van_tol_quasioptical_2005}.   Two samples cut from the same silicon ingot (Samples 1 and 2, see Appendix C for details) were used in this study.
The spin Hamiltonian of an isolated donor is $H_0 = \omega_eS_z + \omega_PI_z^P + AS_zI_z^P$ 
where $\omega_e$, and $\omega_P$ are the Larmor frequencies of the electron and the $^{31}$P nucleus, $S_z$ is the spin angular momentum operator for the electron, $I_z^P$ is the spin operator for the $^{31}$P nucleus and $A$ is the strength of the hyperfine coupling.  The two EPR transitions occur at frequencies $\omega_e \pm A/2$.
Figure~\ref{fig:fig1}(a) shows the echo amplitude decays from  Hahn echo experiments performed on one of the EPR transitions in Sample 1. The figure shows example decays obtained in the dark (blue) and under 1051 nm (E = 1.180 eV) optical excitation (red).
The figure also shows the theoretical prediction (with $\vec{B}_0 \parallel [111]$) from the pair correlation calculation of Witzel {\em et al.}\ \cite{witzel_quantum_2006, witzel_decoherence_2007}(Equation~\ref{eq:Witzel}) which was found to be in excellent agreement with X-band EPR experiments.  The signal is seen to decay much faster than the theory prediction, while optical excitation is seen to increase the observed coherence lifetime. 

The echo intensity observed at time $2\tau$ in an EPR Hahn echo experiment 
\begin{equation}
S(2\tau) = S_0 e^{-2\tau/T_{\text{ID}}}   e^{-2\tau/T_2} e^{-(2\tau/T_{\text{SD}})^n}  \label{eqn:decay}
\end{equation}
typically includes contributions from instantaneous diffusion (ID), the $T_2$ coherence time of the spins, and spectral diffusion (SD) \cite{schweiger_principles_2001}.  Instantaneous diffusion describes the decay observed due to dipolar-induced spin-flips with surrounding electron spins that are also excited by the microwave pulse.   For 10$^{15}$ electron spins/cm$^{3}$ the characteristic time $T_{\text{ID}}$ for instantaneous diffusion is about 1.2 ms \cite{schweiger_principles_2001, tyryshkin_electron_2003} which is significantly longer than the timescales of the decays measured here. Additionally, given the long pulses ($\sim 600$ ns) at 240 GHz, only about one-third of the electron linewidth is excited, further lowering the influence of instantaneous diffusion. 
For isolated donors we would expect the coherence time to be limited by spin-lattice relaxation and $T_2$ = 2$T_1$. In our fits, we ignore the presence of instantaneous diffusion, but correct for the $T_2$ decay using the measured value of $T_1$ (which decreases under optical excitation), before fitting for $T_{\text{SD}}$ and the stretch parameter $n$.  In Si:P, $n$ has been observed to be in the range from 2--3 \cite{tyryshkin_coherence_2006}, while numerical simulations indicate a value of 2.3 \cite{witzel_quantum_2006}.  The values of \TSD~are 109(1), 187(2), and 219(1) $\mu$s for the dark, 1051 nm, 
and Theory curve fits respectively.  Example fits are shown in Appendix E.

\begin{figure*}[htb]
  \includegraphics[width=0.9\linewidth]{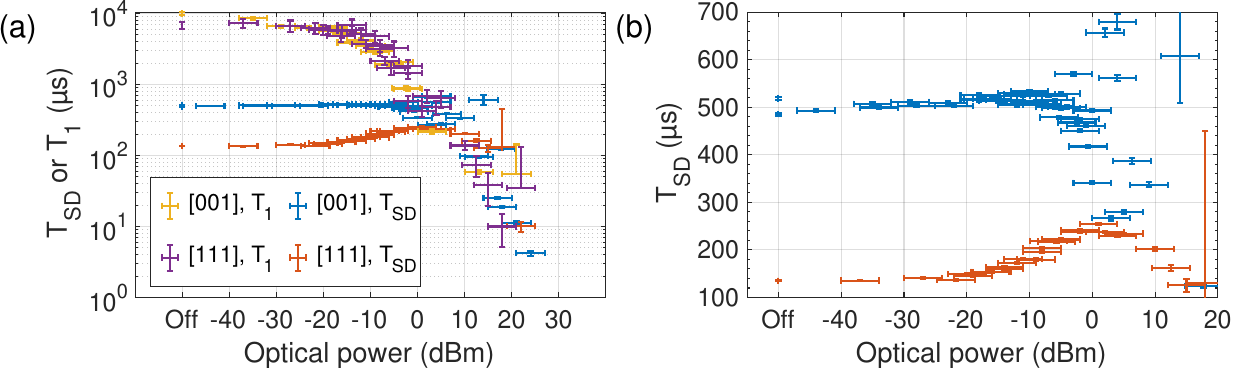}
 \caption{ (a) shows the change in \TSD~and $T_1$ as a function of power of 1047 nm light at both the [001] and [111] crystal orientations in the external field, while (b) zooms in to more clearly illustrate the changes in \TSD~at both orientations.
 Multiple laser configurations were used to create each dataset as described in Appendix C, and the overall uncertainty in laser power indicated by the horizontal error bars is $\pm 3$ dBm. For all plots, the vertical error bars indicate the 95\% confidence interval of the fit. 
 }
 \label{fig:Fig2}
\end{figure*}

Figure~\ref{fig:fig1} (b) shows the measured \TSD~as a function of the optical excitation wavelength for both the EPR transitions.  The optical power used is about 1 mW in these experiments (see Appendix C for experimental details).  The inset shows the CW EPR spectrum of the sample with the two EPR transitions. 
It is clearly seen that the spectral diffusion time increases by approximately 60\% from 110 $\mu$s to about 180 $\mu$s as the energy of the laser is increased above the silicon bandgap ($ 1059.7$ nm at 4.2 K - indirect \cite{green_intrinsic_1990}).   The change in spectral diffusion time is the same at both $^{31}$P hyperfine-split EPR transitions at 8.5875 T (low field, LF) and 8.5915 T (high field, HF). The stretch factor for the decays changes from about 2.1 with below bandgap excitation to about 1.8 with above band-gap excitation (see Appendix E).   The error bars indicate the 95\% percent confidence interval of the \TSD~fit.  The signal measured in these inductively detected pulsed EPR experiments arises from the entire sample. At cryogenic temperatures, the optical penetration depth of light into silicon at wavelengths greater than 1047 nm is known to be several centimeters \cite{macfarlane_fine_1958}, so we expect our samples to be relatively uniformly illuminated.  The wavelength dependence of \TSD~suggests that it is the generation of photo-excited carriers that mediates the increase in the observed \TSD. 



Since higher intensity optical illumination is known to dramatically reduce both $T_1$ and $T_2$ relaxation times for donors in silicon \cite{mccamey_electrically_2012,gumann_nmr_2018} 
we examined how both the $T_1$ and \TSD~times changed with optical power using a fixed 1047 nm wavelength laser with higher maximum power.  
Figures~\ref{fig:Fig2}(a) and (b) show the change in $T_1$ and \TSD~times with laser power at both the [001] and [111] crystal orientations under 1047 nm illumination (in Sample 2).  Similar results were obtained under 980 nm illumination (See Appendix B).  While $T_1$ shows a similar variation with optical power at both crystal orientations, the \TSD~changes with respect to optical power are strongly orientation dependent. 
As the power increases above about 6 dBm, the shortened $T_2$ (resulting from a shortened $T_1$) dominates the echo decays, and a spectral diffusion time cannot be distinguished separately from $T_2$.  The stretch parameter changes from over 2 to 1 in the high power limit indicating a transition to a purely exponential decay (see Appendix E).

In the [001] orientation the \TSD~measured with the laser off is about 500 $\mu$s at 8.59 T, which is about 20\% shorter than the value of 620 $\mu$s at X-band \cite{tyryshkin_coherence_2006}.  At this orientation the \TSD~shows only a very slight enhancement with optical illumination until the $T_1$ times are sufficiently shortened to influence the observed decays.  
In contrast, in the [111] orientation the \TSD~in the dark is about 135 $\mu$s at 8.59 T, significantly shorter than the 270 $\mu$s measured at X-band.  At this orientation \TSD~is seen to {\em increase} as the optical power increases above about -20 dBm, reaching a maximum of 254(6) $\mu$s between 1 and 5 dBm, just as the $T_1$ shortening effects are  becoming apparent.  We estimate the uncertainty in sample orientation to be $\pm 5$ degrees, resulting in a  dispersion of reported \TSD~values that is larger than the error obtained in a single experiment.

\begin{figure}[b]
  \includegraphics[width = 0.9\linewidth]{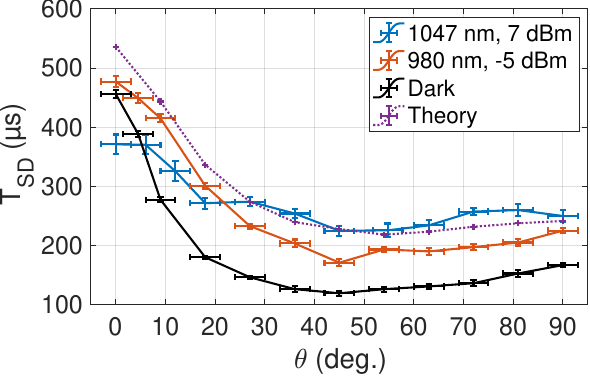}
 \caption{
Characterization of the dependence of the spectral diffusion time \TSD~of the Hahn echo decay fits (Equation \ref{eqn:decay}) as a function of the angular orientation of Sample 2 with respect to the magnetic field. The 90$^\circ$ range of rotation indicated by the horizontal axes of the plots is a rotation about the [1$\bar{1}$0] axis such that 0$^\circ$ is $\vec{B}_0 \parallel [001]$, 54.7$^\circ$ is $\vec{B}_0 \parallel [111]$, and 90$^\circ$ is $\vec{B}_0 \parallel [110]$. 
 }
 \label{fig:Fig3}
\end{figure} 

Figure~\ref{fig:Fig3} shows the variation in \TSD~with crystal orientation in the magnetic field with and without optical excitation. These are compared to \TSD~values predicted by the pairwise correlation calculation (Equation~\ref{eq:Witzel}).  The measurement in the dark at high field shows a significantly larger anisotropy than that predicted by the model.  Defining an anisotropy parameter $\alpha_{\text{ani}} = T_{SD}[001] / T_{SD}[111]$, we see that it changes from a value of about 2 at low field to about 4 at high field (in the dark).
While the effect of optical excitation is small when the crystal is oriented along the [001] direction, \TSD~is enhanced at all other orientations.  In the presence of 980 nm, -5 dBm optical excitation, $\alpha_{\text{ani}}$ is lowered to about 2.8 and is further lowered to 1.7 with 1047 nm, 7 dBm optical excitation. However, at that power, the [001] \TSD~changes are also influenced by the $T_1$ decay (see Figure \ref{fig:Fig2}).
The stretch parameter does not change with orientation (see Appendix E).

It is helpful to examine these experimental results in the context of the central spin model.  The Hamiltonian of the isolated donor interacting with a \Si~spin bath is $H = H_0 + H_{SB}$, where
\begin{multline}
    H_{SB} = S_z \sum_i (A_i I_z^i + B_i I_x^i) + \omega_n \sum_i I_z^i + \sum_{i<j} h_{ij} \ \ \ 
    \label{Eqn:SysHam}
\end{multline}
$A_i$ and $B_i$ are the strengths of the longitudinal and transverse components of the hyperfine interaction with the $i^\textrm{th}$ \Si nucleus. $I_z^i$ and $I_x^i$ are the spin operators for the $i^\textrm{th}$ \Si nucleus. The pairwise nuclear dipolar interaction is $
    h_{ij} = d_{ij}\left(I_x^iI_x^j + I_y^iI_y^j - 2I_z^iI_z^j\right)
$, 
where $d_{ij} = (\gamma_n^2/2r_{ij}^3) \left(3\cos^2\theta_{ij} -1\right)$ depends on the distance between the nuclei and the angle between the inter-nuclear vector and the applied magnetic field.

In Equation~\ref{Eqn:SysHam}, the hyperfine parameter $B_i$ contains dipolar contributions while \(A_i\) contains both contact and dipolar contributions.
The \Si~spins that experience strong hyperfine interactions ($A_i,B_i \gg d_{ij}$) are spectrally separated from the bulk nuclear spin bath and do not contribute to the spectral diffusion process.
At high magnetic field, $B_i \ll \omega_n$, and these hyperfine terms can be dropped in the secular approximation.


After a $\pi/2$ pulse on the central spin, the system is initially in a state of the form $\ket{\psi} = \frac{1}{\sqrt{2}} ( \ket{\uparrow}+\ket{\downarrow})\otimes \ket{J}$, where $\ket{J}$ is a random thermal equilibrium nuclear spin configuration.  The magnitude of the central spin Hahn echo at time \(t = 2\tau\)  is proportional to
\begin{equation}
    \mathcal{S}(2\tau) = \bra{J} e^{iH^-\tau}e^{iH^+\tau}e^{-iH^-\tau}e^{-iH^+\tau}\ket{J}
    \label{eq:echo}
\end{equation} 
where  \(H^{\pm} = \pm \frac{1}{2} \sum_i A_i I_z^i + \sum_{i<j} h_{ij}\)
are different bath Hamiltonians that depend on the
 state of the central spin ($\ket{\uparrow}$ or $\ket{\downarrow}$) \cite{yang_quantum_2016}. 

The echo magnitude in Equation \ref{eq:echo} measures the {\em bipartite entanglement entropy} of the central spin system \cite{niknam_experimental_2021}.  The observed spectral diffusion characterizes the rate of entanglement growth between the central spin and the nuclear spin bath.
The central spin is prepared in close to a pure state  in these experiments  with $\tanh(\hbar\omega_e/2k_BT)\sim 0.89$.  For a mixed thermal state of the nuclear spin bath $\rho = \sum_J p_J \ket{J} \bra{J}$, we average equation~[\ref{eq:echo}] over the different bath states with $p_J \propto \exp(-E_J/k_BT)$.    

At $T$ = 4.2 K and $B_0$ = 8.59 T we have a thermal equilibrium \Si nuclear spin polarization of $4.4 \times 10^{-4}$ which indicates that the nuclear spins are effectively at infinite temperature and all bath configurations are equally likely.  The resulting echo amplitude $
    \mathcal{S}(2\tau) = \text{Tr} \left[e^{iH^-\tau}e^{iH^+\tau}e^{-iH^-\tau}e^{-iH^+\tau}\right]
$
depends only on the electron-nuclear hyperfine and nuclear dipolar couplings, and appears to be independent of the strength of the external magnetic field.
Different approaches have been used to estimate the decay due to this many-body dynamic \cite{yang_quantum_2016}.  For a 2-pulse spin echo experiment the pair correlation function approach has been shown to give very good agreement to experimental results at low magnetic fields \cite{witzel_quantum_2006,witzel_decoherence_2007}.  The resulting echo intensity takes the form
\begin{equation}
    \mathcal{S}(2\tau) = \exp\left(-\sum_{i<k} \frac{d_{ij}^2(A_i-A_j)^2}{4\omega_{ij}^4}\left[\cos \omega_{ij}\tau - 1 \right]^2\right)
    \label{eq:Witzel}
\end{equation}
where $\omega_{ij} = \frac{1}{2} \sqrt{(A_i-A_j)^2 + 4d_{ij}^2}$.  \TSD~is primarily determined by the weakly hyperfine-coupled nuclear spins where the difference in the hyperfine interactions of two nuclear spins is comparable to their nuclear dipolar interactions.  Equation~\ref{eq:Witzel} was used to calculate the theory curves in Figures~\ref{fig:fig1} and \ref{fig:Fig3}.

The sample orientation dependence of \TSD~observed in X-band measurements \cite{tyryshkin_coherence_2006} is primarily due to the anisotropy of the hyperfine interactions $A_i$.  The lattice symmetry significantly influences the donor electron wavefunction within the 1.8 nm Bohr radius of the shallow donor. The strength of the hyperfine interactions is
$
    {A_i} = \frac{16\pi g g_n \beta\beta_n}{3} |\Psi(\bar{r}_i)|^2,
    \label{eq:HfContact}
$
where $g (g_n)$ and $\beta (\beta_n)$ are the g-factor and Bohr-magneton of the electron (nucleus) and the donor wavefunction $\psi(\bar{r})$ can be calculated using an effective mass model originally developed by Kohn and Luttinger (KL) \cite{kohn_hyperfine_1955,kohn_theory_1955}.
While the KL wavefunctions have been shown to qualitatively describe the behavior of the inner shell hyperfine couplings they are not sufficiently accurate to quantitatively predict the strength of the couplings \cite{hale_shallow_1969,hale_shallow_1969-1,ivey_ground-state_1975, ivey_ground-state_1975-1}.  First principles numerical calculations have also shown the failings of the effective mass theory for donors in silicon \cite{swift_first-principles_2020}.  However, the use of KL wavefunctions has yielded excellent quantitative agreement with experimental results at X-band \cite{witzel_quantum_2006,witzel_decoherence_2007}.  

The larger anisotropy in \TSD~seen in the dark in our experiments at high magnetic field suggests that the \Si~hyperfine couplings have a significantly larger anisotropy under these conditions.  This is most likely to arise from a perturbation of the KL wave-functions at high fields. 
In a magnetic field, the donor electron experiences both a Coulomb interaction with the local lattice and a magnetic interaction with the field.  Ignoring the lattice structure of the silicon for the moment, the Hamiltonian of the hydrogenic donor becomes
$ \mathcal{H} = -\frac{1}{2m^*} \left(-i\hbar\vec{\nabla} - e\vec{A}\right)^2 + V(\vec{r})$,
where $m^*$ is the effective mass and $V(\bar{r})$ is the effective central Coulomb interaction.  
We can rewrite the Hamiltonian as $\mathcal{H} = \mathcal{H}_0 + \mathcal{H'}$,
where $\mathcal{H}_0$ is the unperturbed hydrogenic Hamiltonian of the system in the absence of the field and 
$\mathcal{H'} = -\frac{\omega_c}{2}L_z + \frac{1}{8}m^*\omega_c^2(x^2+y^2)$
is the perturbation due to the magnetic field (in the symmetric gauge), where $\omega_c = eB_0/m^*$ is the cyclotron frequency.    
Since $L_z |100\rangle = 0$  for the ground state, (where the eigenstates $|nlm\rangle$ label the principal ($n$), orbital ($l$), and magnetic ($m$) quantum numbers respectively) the first term in this perturbation does not contribute a correction to the ground state wavefunction to the first two orders of perturbation theory. The lowest order correction to the ground state thus arises from the first order perturbation correction due to the second term in the magnetic field perturbation ($\propto B_0^2$). 
Due to the reduced effective mass, the cyclotron energy in silicon is about a factor of 5 times larger than the spin Zeeman energy.
At 8.59 T the cyclotron energy is ~1.2 THz and  $\mathcal{H'}$ can be treated as a perturbation of the $\sim  10.8$ THz Rydberg energy.   The $|200\rangle, \: |300\rangle$ and $|320\rangle$ states all contribute corrections to the ground state wavefunction, expanding the radial size of the wavefunction with the $|320\rangle$ contribution also introducing an anisotropy (see Appendix A).  In the silicon lattice, the symmetry imposed by the applied field will compete with that of the local lattice structure.
Yafet {\em et al}.\ noted that if the magnetic field is increased further, the cyclotron radius of the electron will approach the Bohr radius and eventually become smaller than it \cite{yafet1956hydrogen}, eventually resulting in an increase in the donor ionization energy. A well-known consequence of this is the magnetic-field-induced metal-insulator transition in indium antimonide \cite{keyes1956effect,PhysRevLett.18.773}. 
Keyes suggested that these effects could be important when silicon donor qubits are used for quantum technologies \cite{keyes2005effects}.

Given the delocalized nature of the donor wavefunction, even a small change in spatial extent due to the magnetic field can influence the hyperfine interactions of a large number of \Si~bath spins.  An increase in the spatial extent of the donor wavefunction leads to a larger number of nuclear spins contributing to the spectral diffusion process, and a faster decay of the Hahn echo. We propose that this is the origin of both the shortened \TSD~and the increased anisotropy with crystal orientation in the field.  The increase in \TSD~under low-power optical excitation suggests that the cyclotron motion of the electrons is disrupted by the light -- potentially due to interactions with excited conduction band electrons. As a result the measured \TSD~values are closer to their low-field values which also results in a smaller anisotropy of \TSD~under optical excitation.

An increase in \TSD~under optical excitation could potentially also arise if the \Si~spins are hyperpolarized during the optical excitation.  For NV centers in diamond, polarizing the spin bath (electronic P1 centers) has been shown to suppress spectral diffusion and increase \TSD~\cite{takahashi_quenching_2008} 
While above bandgap illumination has been shown to hyperpolarize the phosphorus donor spins at high field and low temperature \cite{mccamey_fast_2009,gumann_nmr_2018}, only weak optical hyperpolarization of the silicon spins has been observed to date at high magnetic fields \cite{verhulst_optical_2005}. Additionally, the dark curve in Figure~\ref{fig:fig1}(a) was collected immediately (on the order of a minute) after the above band gap excitation was turned off, showing that the enhancement effect switches on and off much faster than the time scale of relaxation of \Si nuclear $T_1$ (on the order of a few hours \cite{verhulst_optical_2005}).
We ran a series of echo-detected field-swept EPR experiments in the presence of optical excitation and did not see any evidence of phosphorus or silicon nuclear spin hyperpolarization (see Appendix B).   We did observe a shift in the mean position of the two peaks, likely a direct result of the magnetic susceptibility of the conduction band electrons in silicon.  

In conclusion, we showed that \TSD~of phosphorus donors due to the adjacent \Si~spins measured at high magnetic fields are shorter than those measured at lower magnetic fields and have a larger anisotropy with respect to crystal orientation in the field.  We suggest that this is due to a perturbation of the ground state electronic orbital of the donor electron due to the magnetic field.  The presence of low-power optical excitation is seen to both increase \TSD~and reduce the anisotropy, approaching the values measured at low magnetic fields.  Such magnetic-field-induced changes to the donor wavefunction could influence the control of donor qubits.

\noindent {\em Acknowledgements:} 
We thank Rufus Boyack, William Coish, James Logan, Brendan Sheehan, Susumu Takahashi and Bhargava Thyagarajan for helpful discussions.  This work was supported in part by the National Science Foundation under grants OIA-1921199.  Part of the work was done at the National High Magnetic Field Laboratory, which is supported by the NSF under grant DMR-1644779, DMR-2128556 and the State of Florida.  We acknowledge the support of a QISE-NET Triplets Award with EQW funded by NSF award DMR-1747426.

\setcounter{figure}{0}         
\renewcommand\thefigure{A\arabic{figure}} 

\bibliography{SiPRefs}
\newpage
\vspace*{-0.2in}
\section{Appendix A: Ground State of Hydrogenic Donor in a Strong Magnetic Field}
\vspace*{-0.1in}

 In the presence of a magnetic field electrons undergo cyclotron motion.  Typically the magnetic energy is much lower than the Coulomb energy so
the energy levels of the hydrogen atom are only corrected to lowest order in perturbation theory.  The dominant correction is due to the \(L_z\)
angular momentum term about the magnetic field (linear Zeeman effect). This shifts the energies of the states with non-zero orbital angular momentum
(p-states and higher with $m\neq $0) as studied previously \cite{murdin_sip_2013}. The only correction to the ground state arises from the quadratic Zeeman correction.
The perturbation Hamiltonian takes the form
$$H' = \frac{e}{2m^*}\left(\bar{p}\cdot\bar{A} + \bar{A}\cdot\bar{p}\right) + \frac{e^2}{2m^*}A^2.$$

\noindent We choose the magnetic field to be along the $\hat{z}$ direction. Choose a symmetric
gauge with $A_x = -\frac{1}{2}yB_0$ and $A_y = \frac{1}{2}xB_0$. We can then write the perturbed Hamiltonian in the following form 
\begin{eqnarray}
 H' & = & \frac{ie{\hbar}B}{2m^*}\left(x\frac{\partial}{\partial y} - y\frac{\partial}{\partial x}\right) + \frac{e^2B^2}{8m^*}\left(x^2 + y^2\right) \nonumber \\
 & = & -\frac{eB}{2m^*}L_z + \frac{e^2B^2}{8m^*}\left(x^2 + y^2\right) \nonumber.
 \end{eqnarray}

\noindent Using the standard notation $\psi_{nlm}$, where $n,l,m$ correspond to the principal, orbital and magnetic quantum numbers respectively, the unperturbed ground state wavefunction is $$\psi_{100}(r,\theta,\phi) = \langle r,\theta,\phi | \psi_{100}\rangle = \frac{1}{\sqrt{\pi}} \left(\frac{1}{a_ 0}\right)^{3/2} e^{-r/a_0}$$
\noindent where
$$a_0 = \frac{4\pi \epsilon \epsilon_0 h^2}{m^*(2\pi e)^2} = \frac{ \
\epsilon h^2}{k m^*(2\pi e)^2} = \frac{\epsilon}{m^*/m_e} a_H.$$
\noindent The lowest order of perturbation theory tells us that the corrections to the ground state wavefunction take the form
$$|\psi_{100}^{(1)}\rangle = \sum_{k>1} \sum_{l,m} \
\frac{\langle klm|H'|100\rangle} {E_k - E_ 1} |klm\rangle$$
\noindent where
$$E_k = -\frac{1}{2k^2} \left( \frac{1}{4 \pi \epsilon \epsilon_0} \
\frac{e^2}{a_ 0}\right)  = E_k^H \frac{m^*}{\epsilon^2 m_e}.$$

\noindent Since $L_z |100\rangle = 0$, the only term in the perturbation that matters is the quadratic term \(B^2\). { }The term $x^2 + y^2$ can be rewritten as $r^2 \sin^2\theta$. { }The integration
is over $d^3r \equiv r^2 \sin \theta dr d\theta d\phi$.

We can examine the overlap with the first few excited states ($n = 2,3$).  It is easy to see that the terms $\langle klm|H'|100\rangle = 0$ for any $m \neq 0$.  In addition the terms $\langle 210|H'|100\rangle$ and $\langle 310|H'|100\rangle$ both go to zero.  The only non-zero contributions are due to the $|200\rangle$, $|300\rangle$ and $|320\rangle$.  We have
$$\psi_{200}(r,\theta,\phi)= \frac{1}{\sqrt{\pi}} \left(\frac{1}{2a_ 0\
}\right)^{3/2} \left(1 - \frac{r}{2a_ 0}\right)e^{-r/2a_ 0}$$

\noindent with $E_2 = 1/4E_1$ and $\frac{1}{E_2 - E_1} = -\frac{4}{3E_1}$;
\begin{eqnarray}
\psi_{300}(r,\theta,\phi) & = &\frac{1}{\sqrt{\pi}} \left(\frac{1}{3a_ 0\
}\right)^{3/2}  \times \nonumber \\
& & \hspace*{0.2in} \left[ 1 - \frac{2r}{3a_ 0} + \
\frac{2}{27}\left(\frac{r}{a_ 0}\right)^2\right] e^{-r/3a_ 0};\nonumber 
\end{eqnarray}
\noindent  and 
\begin{eqnarray} \psi_{320}(r,\theta,\phi) & = & \frac{1}{3\sqrt{2\pi}} \
\left(\frac{1}{3a_ 0}\right)^{3/2}  \times \nonumber \\
& & \hspace*{0.3in}  \left(\frac{r}{3a_ 0}\right)^2 e^{-r/3a_ 0}\left(3 \cos^2\theta - 1\right) \nonumber 
\end{eqnarray}

\noindent with $E_3 = E_1/9$ and $\frac{1}{E_ 3-E_ 1} = -\frac{9}{8E_ 1}$.  

At 8.6 T, using $m^* = m/5$ and $\epsilon = 11.7$ for silicon, we get $a_0 = 3.1$ nm.  Evaluating the overlap integrals in Mathmatica results in the following normalized ket for the first order perturbed ground state.
$$|100\rangle^{(1)}  = 0.995|100\rangle + 0.100|200\rangle + 0.006|300\rangle + 0.017|320\rangle$$
The contributions from the $|200\rangle$ and $|300\rangle$ states expand the radial extent of the wavefunction while the $|320\rangle$ term also breaks the spherical symmetry of the unperturbed wavefunction.


\begin{widetext}
\begin{figure*}[htb]
    \includegraphics[width=0.8\linewidth]{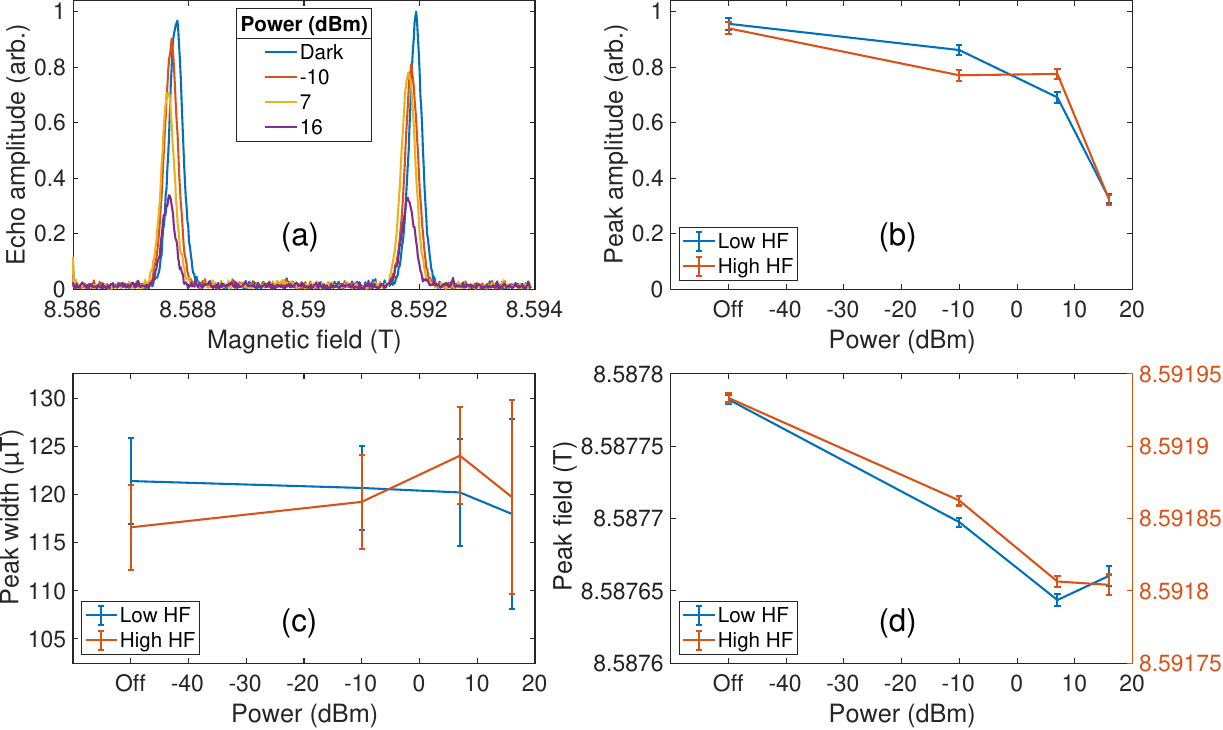}
    \caption{Echo-detected field sweeps with 1047 nm optical excitation.}
    \label{fig:EDFS}
\end{figure*}
\end{widetext}

\section{Appendix B: Additional Data}

\subsection{Echo-Detected Field Sweeps under Optical Excitation} Figure~\ref{fig:EDFS}(a) shows the results of a series of  echo-detected field-sweep EPR experiments ($\tau = 8 \mu$s) when the sample is irradiated with a 1047 nm laser at different optical power levels.  Figure~\ref{fig:EDFS}(b), (c) and (d) show how the amplitude, width and mean position of the two EPR peaks change with laser excitation.  The simultaneous drop in amplitude and shift in the mean peak positions suggests that these are due to the presence of photoexcited carriers. The drop in amplitude is consistent with the shortened T$_1$ relaxation time,
while the shift in the mean position of the two peaks (on the order of 1 part in $10^4$) is likely a direct result of the changing magnetic susceptibility of the conduction band electrons in the silicon. As the electron density ($n$) increases with applied laser power both T$_1$ and  $\chi \sim 3n \mu_0 \mu_B^2 / 2 E_F$ \cite{simon_oxford_2013} change.
Note that the width of the peaks does not change with laser excitation, indicating that there is no hyperpolarization of the \Si~spins at the laser intensities and durations studied. The shift in the resonance peaks also suggests that donor ionization is not responsible for the drop in the EPR signal, at least at low laser power.

\subsection{Electron-Nuclear Double Resonance}


\begin{figure}[ht]
  \includegraphics[width=0.85\linewidth]{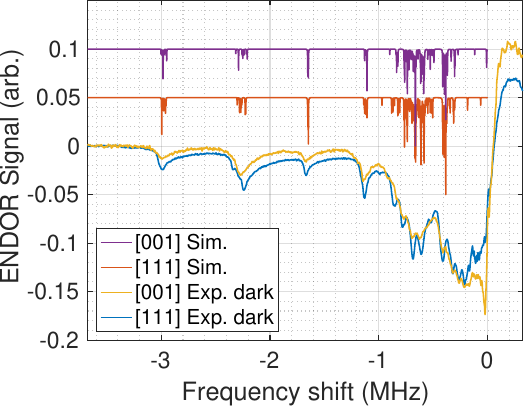}
  \caption{ 
 Comparison of ENDOR spectra acquired on Sample 2 at 8.592 T.  The Frequency axis indicates the deviation of the RF drive from the \Si Larmor frequency, which is 72.7 MHz. (a) shows the experimentally acquired spectra without optical excitation (``Exp. dark'') for the [111] and [001] crystal orientations. The simulated spectra labeled ``Sim.'' are calculated from hyperfine parameters reported by Ivey and Mieher in \cite{ivey_ground-state_1975}. 
}
 \label{fig:Fig5_ENDOR}
\end{figure}

\noindent Electron-nuclear double resonance (ENDOR) experiments on donors in silicon were first performed by Feher \cite{feher_electron_1959, feher_electron_1959-1} at X-band. Ivey, Hale and Mieher calculated and further characterized theoretically and experimentally more hyperfine couplings between the phosphorus electron and the more weakly coupled \Si nuclei \cite{hale_shallow_1969, hale_shallow_1969-1, ivey_ground-state_1975, ivey_ground-state_1975-1} that would contribute to spectral diffusion. We performed ENDOR experiments on Sample 2 to examine the effects of crystal orientation and optical excitation on the hyperfine parameters.

In our experiment, we monitor the continuous wave EPR signal of the higher field (8.592 T) $^{31}$P hyperfine transition while an RF field is swept with increasing frequency over the range of \Si nuclear resonance frequencies, which are shifted from the nominal Larmor frequency (72.7 MHz) by their hyperfine interaction with the donor electron spin. The RF field is applied with a coil that is wound 6 times around the sample mount. A modulation coil for the lock-in detection of the continuous wave EPR signal is positioned below the sample and produces an estimated 3 $\mu$T amplitude modulation to the $B_0$ field. The RF is swept at a rate of 13 kHz/s.


In Figure \ref{fig:Fig5_ENDOR}, the ENDOR spectra collected in the dark are compared for the [001] and [111] orientations. Only one half of the full spectra are shown to better visualize the comparison to theory.
Simulations of the spectra are computed based on the experimental hyperfine parameters determined by Ivey and Mieher at X-band \cite{ivey_ground-state_1975}. 
The simulations broadly recreate the spectra well for couplings stronger than 1 MHz in magnitude.   However, these strong hyperfine do not contribute to the spectral diffusion process as they are significantly stronger than the nuclear spin dipolar couplings on the order of a few hundred hertz at most -- which are captured by the broad featureless peak in the center.

\subsection{Spectral Diffusion under 980 nm illumination}

\begin{figure}[htb]
  \includegraphics[width=0.9\linewidth]{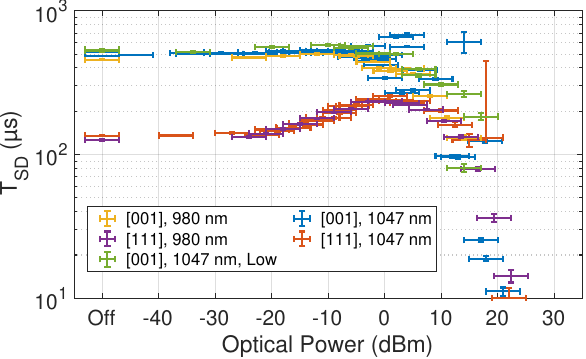}
    \caption{Change in \TSD~and $T_1$ as a function of power of 980 nm light at both the [001] and [111] crystal orientations in the external field.  Comparison to 1047 nm excitation and experiments performed on the LF transition.}
    \label{fig:power980}
\end{figure}

Figure~\ref{fig:power980} shows the \TSD~and $T_1$ behavior as a function of laser power at 980 nm.  The behavior is seen to be almost identical to that observed at 1047 nm (Figure~2 in the main paper).  The same behavior is observed for experiments performed on both the LF and the HF transitions.  At 980 nm the penetration depth reduces to about 100 $\mu$m  so we expect the illumination to be fairly inhomogeneous. 
``Theory'' curves are from pairwise correlation simulations as described in Appendix D.

\section{Appendix C: Experimental Details}
\label{Apdx:ExpDetails}

\subsection{High Field Pulsed EPR}

\begin{figure*}[ht]
	\begin{center}
   		 \includegraphics[width=0.9\linewidth]{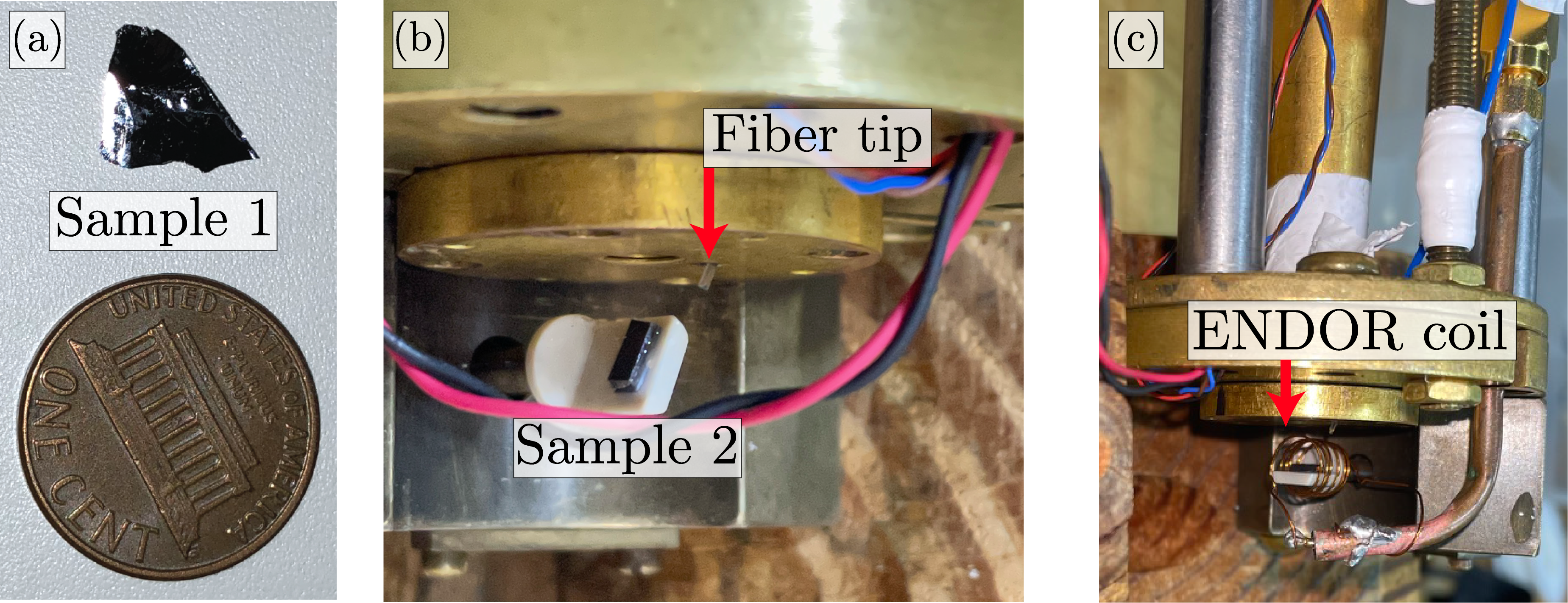}
 		 \caption{\label{fig:s2} (a) Picture of Sample 1 with penny for scale. (b) Sample 2 is seen on the rotating platform at the end of the 240 GHz EPR probe in roughly the [111] orientation. The bare tip of the optical fiber protrudes about 1 mm from the flange bore. Left of the fiber tip and above the sample is the hole at the end of the corrugated waveguide. When the probe is in the magnet, the $\vec{B}_0$ field is upward and the $\vec{B}_1$ field is horizontal. (c) The sample is in the [001] orientation with the RF coil for performing ENDOR wrapped around it. The optical fiber with the blue coating can also be seen in this view, and it is tightly taped to a support screw to fix it in place. The $B_0$ modulation coil is not pictured, but is positioned a few mm below the sample in continuous wave EPR and ENDOR experiments.}
         \label{fig:FigAux_SamplePics}
	\end{center}
\end{figure*}

\begin{figure}[hb]
	\begin{center}
   		 \includegraphics[width=.5\textwidth]{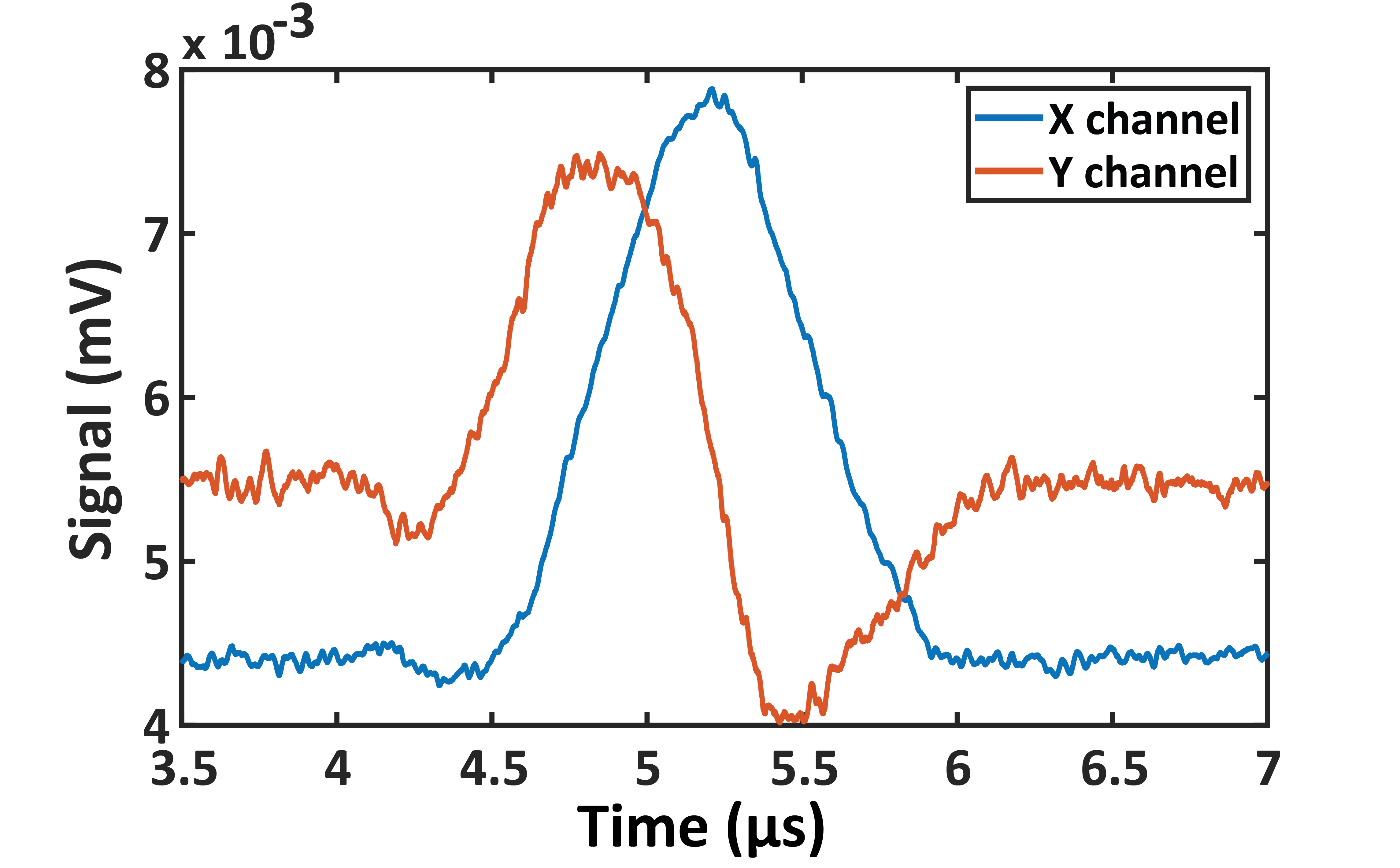}
 		 \caption{The in-phase and quadrature echo signals from a single-shot Hahn echo experiment performed on the low field transition on the bulk sample at 5 K in the dark.}
      \label{fig:echoes}
	\end{center}
\end{figure}

\noindent The high field EPR experiments were performed at the National High Magnetic Field Laboratory (NHMFL) in Tallahassee, Florida on the 240 GHz electron magnetic resonance setup \cite{van_tol_quasioptical_2005,morley2008multifrequency}. 
This instrument can run both continuous wave and pulsed EPR experiments at temperatures as low as to 2 K. The 240 GHz microwaves are delivered to the sample using a quasi-optical design to optimize sensitivity and minimize losses \cite{earle_farinfrared_1996}. The design has been adapted to allow for phase sensitive detection \cite{morley2008multifrequency}.

The setup has been adapted to allow for optical excitation of the sample. A multi-mode optical fiber (ThorLabs FT600EMT) is fed through the probe with the bare tip directed at the sample and positioned about 6 mm away as shown in Figure~\ref{fig:FigAux_SamplePics} (b) and (c).


Figure~\ref{fig:echoes} shows an example detection of a single-shot echo signal from the lower hyperfine transition on Sample 1 at 5 K in the dark.  Both the in-phase and out-of-phase channels are shown. The time interval ($\tau$) between the first and second microwave pulse is 2.5 $\mu$s.  The phase stability of the spectrometer is not optimal over a long period of time due to the thermal shrinkage and expansion over the long microwave transmission and receiving pathway. The phase can be adjusted by a phase shifter. This phase instability does not affect the intensity of the total echo signal, calculated by integrating the signal in both channels and adding them together in quadrature. 

\subsection{Samples}
Two samples have been examined in this work. Both samples originated from the same ingot and have the same donor concentration. Sample 1 and Sample 2 are shown in Figure \ref{fig:FigAux_SamplePics}(a) and (b) respectively.

Sample 1 is a larger rough-cut sample with approximate dimensions $10 \times 5 \times 3 \text{ mm}^3$. In experiments performed on Sample 1, the orientation of the crystal lattice with respect to the magnetic field is not precisely known. These experiments were performed with the sample on a stationary platform. Based on comparisons to later results, it is clear that for Sample 1, the magnetic field was not aligned with [001] (or any permutation thereof), and was likely close to a [111] or [110]- type orientation. This is evidenced by the fact that for [001]-type orientations, the \TSD~is nominally longer than it is for any other orientation, and there is little to no optical enhancement of \TSD, which is not the case for the experiments on Sample 1. Sample 1 was used in the initial discovery of the optically induced \TSD~enhancement.

Sample 2 is smaller with dimensions $4.3 \times 2.2 \times 1.1 \text{ mm}^3$. To prepare Sample 2, a 1.1 mm thick section was cut from the source ingot (with [001] growth axis) using a circular saw blade on a precision cutter, exposing the [001] plane. Then, using a diamond wire saw, a rectangular portion of the thin section was cut out. The crystal orientation of Sample 2 was determined by a series of Hahn echo experiments. The sample was first placed on a rotatable mount such that it could be rotated about the [001] axis, which pointed perpendicular to the $B_0$ magnetic field. This step was performed on a different rotating platform than the one in Figures \ref{fig:FigAux_SamplePics} (b) and (c). The orientation for which \TSD~was maximum was recognized as the [100] direction. Identification of this direction enabled the identification of the [110] axis as 45$^\circ$ from [100] and 90$^\circ$ from [001]. For the experiments presented in the main text, Sample 2 was placed on a second rotating platform (shown in Figures \ref{fig:FigAux_SamplePics} (b) and (c)) such that the plane in which the sample lay coincided with the axis of rotation. In this manner, we were able to rotate the sample continuously through a 90$^\circ$ range such that $\vec{B}_0$ passed through [001], [111], and [110].




\subsection{CW EPR}
The spectra in Figures~1 and 4 in the main paper were obtained under fast-passage conditions with a 40 kHz modulation frequency and 0.1 mT modulation strength. The microwave power used was 20 $\mu$W.

\subsection{Optics}
Light was delivered to the sample via a 3.7 m optical fiber running from the optics table to the magnet and down the length of the probe. Four lasers were used in the experiment.

\begin{figure}[ht]
  \includegraphics[width=\linewidth]{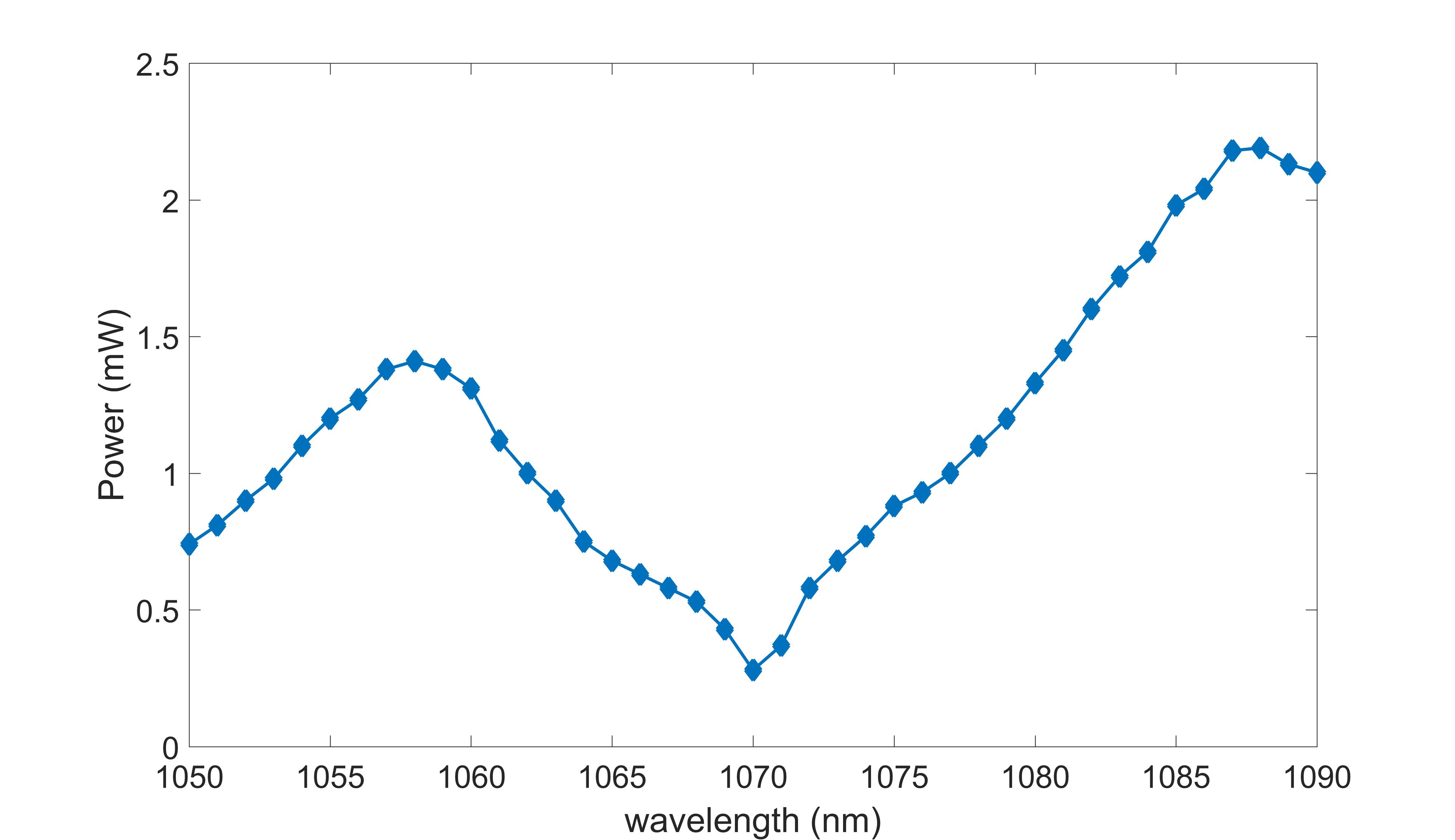}
  \caption{ Power characterization of the O/E Land OETLS-300-1060 tunable laser.
 }
 \label{fig:power_lambda}
\end{figure}

The 980 nm laser diode was mounted in a ThorLabs diode controller. Various configurations were used to partially block the light entering into the fiber from the 980 nm laser diode. The 1047 nm laser is a CivilLaser 100 mW fiber-coupled laser. The tunable laser is a O/E Land OETLS-300-1060 electrically-tunable external-cavity fiber laser, for which the power vs. wavelength is plotted in Figure \ref{fig:power_lambda}.

The optical phenomena are seen to take effect over a range of optical powers between roughly -35 dBm and +25 dBm. This 60 dB range is greater than the range of power that our lasers can output. Accordingly, multiple configurations of the optics were used to block varying amounts of light from reaching the sample. For experiments with the 980 nm laser, once a configuration was set, the power was adjusted over a 20 dB range by varying the diode current. The ranges of powers for the different configurations partially overlapped, and by matching the overlapping segments of collected data, the overall attenuation factor for each configuration was determined, enabling the data to be represented over the full $60$ dB range of power.

For experiments with the 1047 nm laser (performed at a later date, after the 980 nm laser experiments), again multiple configurations were used, but the diode was kept at constant current in each configuration, and the power was adjusted with a pair of fiber-coupled variable attenuators (Thorlabs VOAMMF). A 10/90 power splitter (Thorlabs TM200R2F1B) with one output directed to a power meter (Thorlabs PM20A) enabled more precise calibration and monitoring of the power in each experiment. These results matched well with the results from the 980 nm laser experiments, confirming the effectiveness of those methods.

There are a number of effects that contribute to the uncertainty of the effective power reaching the sample. We believe that the most significant source of uncertainty in the reported optical power arises due to the different angular orientations of the sample with respect to the emitted beam of light. The amount of optical penetration should depend strongly on the cross-sectional area of the sample within the beam, and the angle of incidence of the light. The rotating sample holder itself also diffuses the infrared light to varying degrees depending on the orientation in a given experiment. We estimate the uncertainty in the power to be $\pm 3$ dBm. These uncertainties do not detract from our results as \TSD~and $T_1$ do not sharply depend on wavelength or power. The powers in the reported data indicate the calibrated measurement of the power entering into the 3.7 m optical fiber that guides light from the optics table to the sample.

\subsection{Data Collection}
In the experiments on Sample 1, pulse durations were 500 ns and 950 ns for the $\pi/2$ and $\pi$ pulses respectively. The repetition time of the experiment was 100 ms, and 16 repetitions were performed for signal averaging.

In the experiments on Sample 2, pulse durations were 700 ns and 900 ns for the $\pi/2$ and $\pi$ pulses respectively. In the Hahn echo experiments, the repetition time of the experiment was 20 ms, and 50 repetitions were performed for signal averaging. The $T_1$ measurements were acquired with inversion recovery experiments, $\pi - \tau_1 - \pi/2 - \tau - \pi - \tau - \mathrm{echo}$, where $\tau_1$ is varied. The first $\pi$ pulse is toggled on and off, and the difference in echo magnitudes decays as $\exp[-\tau_1/T_1]$. In inversion recovery experiments, the repetition time was set to 50 ms, and 25 repetitions were performed for signal averaging.

We have not strictly controlled for time-dependent effects of the optical excitation with regard to build-up and cool-down times of applying the laser light, however we see that generally the toggling on or off of the enhancement effect is immediate on the timescale of experimental operation. Hahn echo experiment durations vary depending on the number of averages, repetition time, and number of $\tau$ values sampled, but generally each experiment takes 10 to 20 minutes to complete. The typical downtime between experiments is about two minutes. In the 1047 nm data shown in Figure 3 of the main paper, the experiments were performed by alternating between adjusting the sample orientation and turning on/off the laser as follows:\newline

\noindent -- set orientation 1, acquire w/o laser, acquire w/ laser \newline
-- set orientation 2, acquire w/ laser, acquire w/o laser \newline
-- set orientation 3, acquire w/o laser, acquire w/ laser \newline
-- set orientation 4, acquire w/ laser, acquire w/o laser \newline
-- repeat sequence \newline

The contrast between the laser on and laser off data sets is evidence of the toggling timescale of the enhancement effect being on the order of a minute or shorter.


\section{Appendix D: Numerical Simulations}
\label{Apdx:NumericalSims}

\begin{widetext}
\begin{figure*}[htb]
  \includegraphics[width=0.7\linewidth]{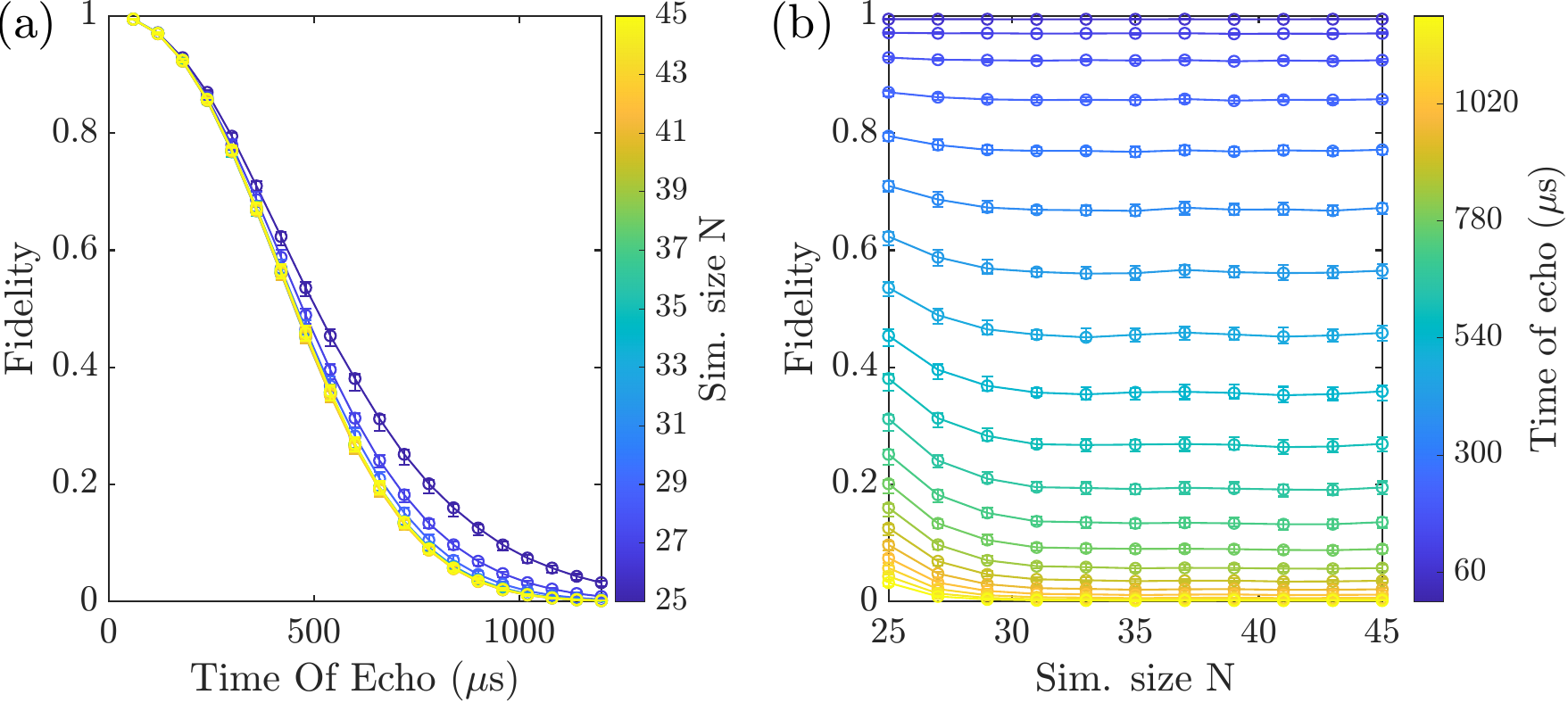}
  \caption{Convergence test for pairwise correlation simulations of Hahn echo decay. (a) and (b) show the same data with the horizontal and color axes exchanged. (a) Hahn echo decay fidelities for various simulation sizes. (b) Fidelities obtained with different simulation sizes.
 }
 \label{fig:FigAux_SimNConverge}
\end{figure*}
\end{widetext}

\noindent To obtain the ``Theory'' values \TSD~as a function of angle in Figure 3 as well as the ``Theory'' curves in Figure 1 of the main paper, we ran a set of simulations based on the pairwise correlation calculation of Reference \cite{witzel_quantum_2005}. The simulation generates a silicon crystal lattice that is a sphere of diameter $Na_0$, where $a_0 = 5.43$ \r{A} is the silicon unit cell width. The central site is the location of the $^{31}$P donor. The other lattice sites are randomly seeded with $^{29}$Si spins using a binomial random number generator with a site occupation probability of 4.7\%. Hyperfine couplings between each \Si and the donor electron are calculated in the method of Reference \cite{de_sousa_electron_2003} (Equation 57). For sites within the Bohr radius of the donor electron, the hyperfine coupling is calculated as the contact interaction. For sites outside the Bohr radius, the dipolar coupling is added. Then the \Si nuclear dipolar couplings are calculated for all pairings of \Si nuclei. The fidelity, or signal $\mathcal{S}$, is calculated with Equation 4 in the main paper. Fidelities are computed for 48 random realizations of the lattice. In Figures 1(d) of the main paper and \ref{fig:FigAux_SimNConverge} 
the  ``Theory'' value is the median fidelity from the 48 realizations, and the errorbars are the boundaries of the top and bottom quartiles.

We calculated the decays of a [001] orientation crystal simulation as a function of lattice size $N$ and found that the results converged with discrepancies $<0.001$ when $N \gtrsim 39$ such that the average number of $^{29}$Si spins was $<12000$, in agreement with the results from Ma \textit{et al.} \cite{ma_uncovering_2014}. The results of this convergence test are shown in Figure \ref{fig:FigAux_SimNConverge}. The simulation with $N=39$ was repeated for rotations of the lattice at multiples of 9$^\circ$ from 0$^\circ$ to 90$^\circ$. To obtain the values of \TSD~and $n$ for the theory curves in Figure 3 of the main paper, each fidelity curve is fit to a stretched exponential of the form $e^{-(2\tau/T_\mathrm{SD})^n}$.

\section{Appendix E: Data Analysis}

Signal magnitudes are acquired by integrating over the echo signal obtained with an oscilloscope as seen in Figure \ref{fig:echoes}. This records a pair of values, corresponding to the in-phase $I$ and quadrature $Q$ components of the signal. The signal magnitude then is $\sqrt{I^2 + Q^2}$. In fitting the data, we ignore instantaneous diffusion as it is expected to cause decay on the timescale of 1.2 ms, which is sufficiently longer than the nominal spectral diffusion. We therefore fit the magnitudes of the Hahn echoes to the function
\begin{equation}
    Ae^{-(2\tau/T_2) - (2\tau/T_\mathrm{SD})^n} + B
\end{equation}
where the free parameters are $A$, $T_\mathrm{SD}$, $n$ and $B$. The value of $T_2$ is $2T_1$, which is determined by identifying the inversion recovery experiment which was performed with conditions most closely resembling the Hahn echo experiment. The parameter $B$ accounts for offsets due to a non-zero noise floor of the echo magnitudes.

\subsection{Example Data Fits}

Here are examples of the echo decay and inversion recovery experiment signals for Figures 1--3 in the main paper. Each signal is normalized, and its offset is subtracted using its fit parameters.  The stretch factors for Figures 2 and 3 are shown in the following section. ``Theory'' curves are from pairwise correlation simulations as described in the Appendix D above.

\begin{table*}[t]
    \centering
    \label{tab:figures_table1}
    \begin{tabular}{ p{6cm}  p{6cm} }
        \includegraphics[width=\fitFigScale\linewidth]{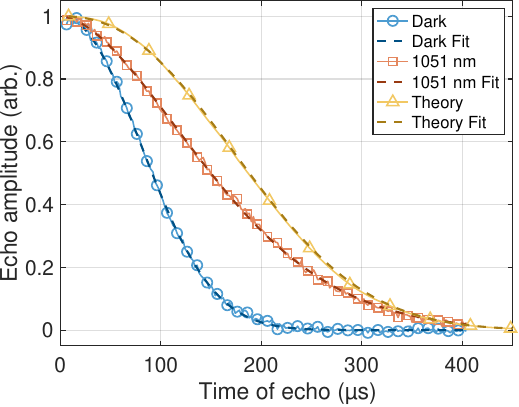} 
        &
        \includegraphics[width=\fitFigScale\linewidth]{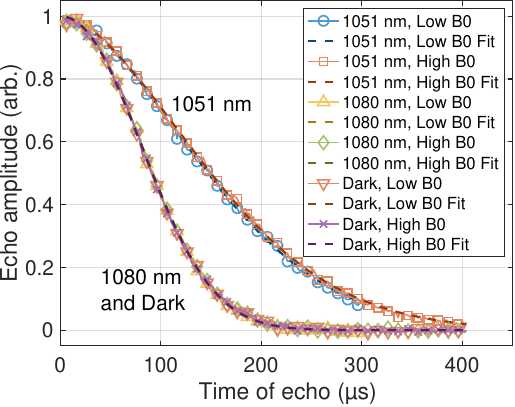} 
    \end{tabular}
        \caption{
        Example fits for (left) Fig.~1a: Hahn echo, Sample 1. (right) Fig.~1b: Hahn echo, Sample 1. All 1051 nm lines are closely overlapped, and all 1080 nm and Dark lines are overlapped as well. The two groupings are labeled accordingly.}
\end{table*}

\begin{table*}[t]
    \centering
    \label{tab:figures_table2}
    \begin{tabular}{ p{6cm}  p{6cm} }
        \includegraphics[width=\fitFigScale\linewidth]{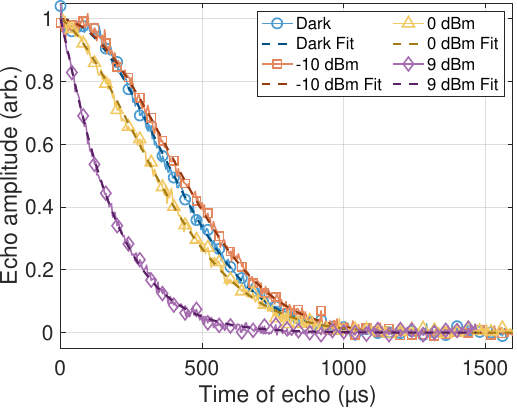} 
        &
        \includegraphics[width=\fitFigScale\linewidth]{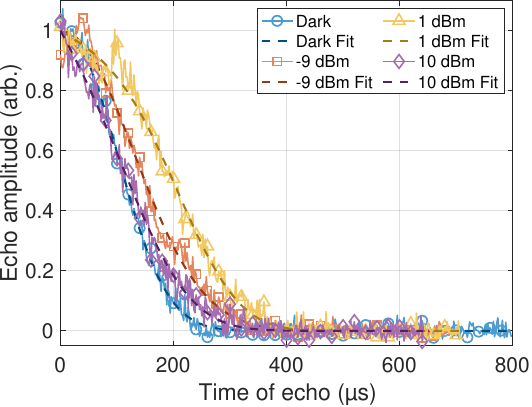} \\ 
        \includegraphics[width=\fitFigScale\linewidth]{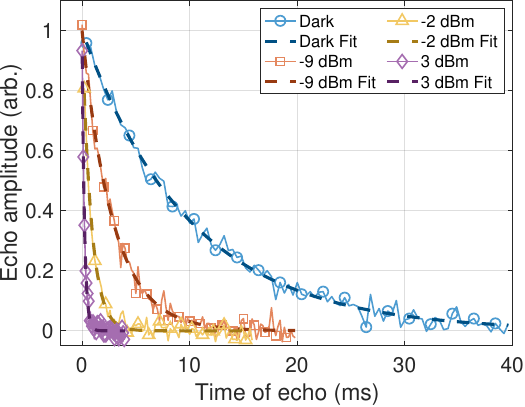} 
        &
        \includegraphics[width=\fitFigScale\linewidth]{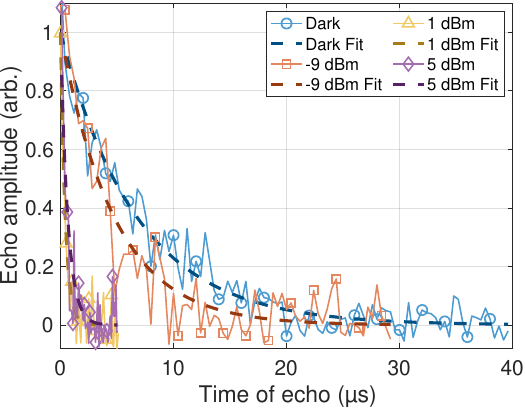} \\ 
        \end{tabular}
        \caption{Examples for Figure~2: (top left) Hahn echo, Sample 2, [001] alignment;
        (top right) Hahn echo, Sample 2, [111] alignment; 
        (bottom left) Inversion recovery, Sample 2, [001] alignment; 
         (bottom right) Inversion recovery, Sample 2, [111] alignment.}

\end{table*}

\begin{widetext}

\begin{table*}[t]
    \centering
    \label{tab:figures_table3}
    \begin{tabular}{ p{6cm}   p{6cm}  p{6cm}  }
        \includegraphics[width=\fitFigScale\linewidth]{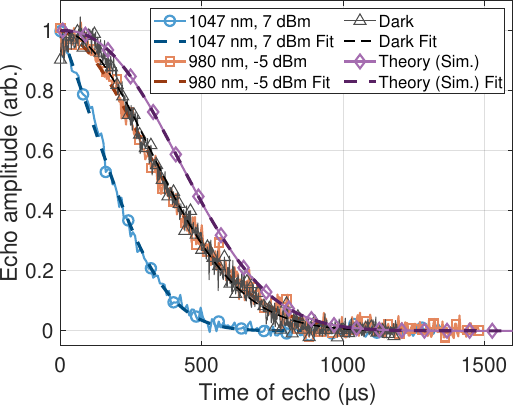}  & 
        \includegraphics[width=\fitFigScale\linewidth]{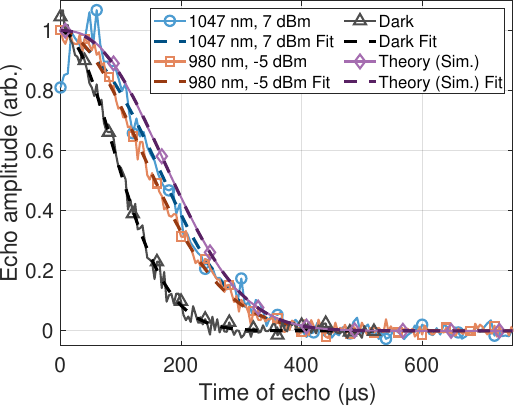} & 
        \includegraphics[width=\fitFigScale\linewidth]{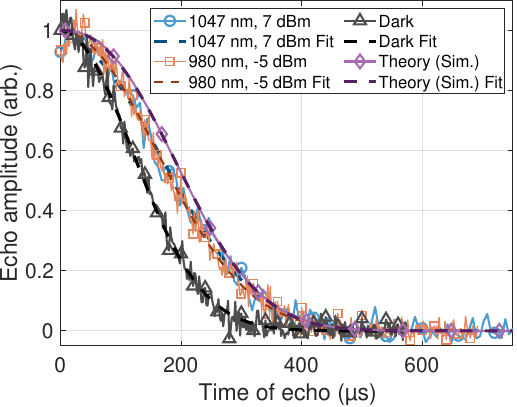} \\   

        
    \end{tabular}
         \caption{Example fits for Figure 3: (left) Hahn echo, Sample 2, angle 0 deg. (also [001] alignment). Here the 980 nm and Dark lines are closely overlapped.
        (middle)  Hahn echo, Sample 2, angle 55 deg. (also [111] alignment). (right)
        Hahn echo, Sample 2, angle 90 deg. (also [011] alignment).}
\end{table*}
\end{widetext}

\clearpage

\subsection{Stretch Factors for Fits}

Here we plot the stretch factors $n$ corresponding to the \TSD~values reported in Figures~2 and 3 of the main paper.

\begin{figure}[htb]
  \includegraphics[width=0.9\linewidth]{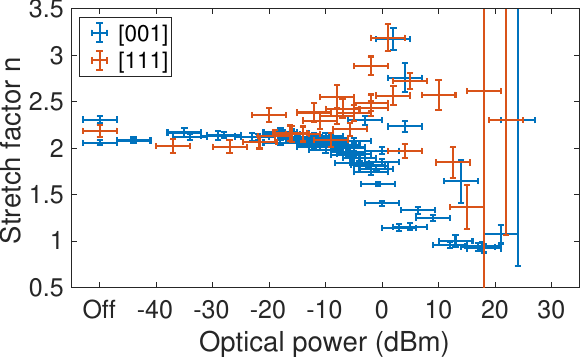}
    \caption{Change in exponential decay stretch factor $n$ as a function of illumination power with 1047 nm light.}
    \label{fig:power_dep_n}
\end{figure}

\begin{figure}[htb]
  \includegraphics[width=0.9\linewidth]{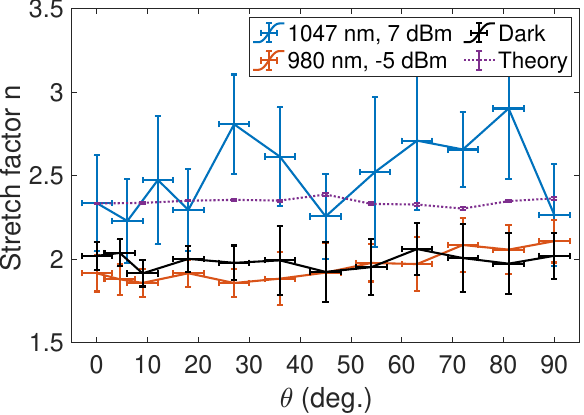}
    \caption{The exponential decay stretch factor $n$ as a function of angle.}
    \label{fig:angle_dep_n}
\end{figure}


\end{document}